\documentclass[manuscript]{aastex}

\slugcomment{}

\shorttitle{$f(R)$, $f(T)$ and $f(\mathcal{G})$ models by variable deceleration parameter }
\shortauthors{F. Darabi}

\begin{document}


\title{Reconstruction of $f(R)$, $f(T)$ and $f(\mathcal{G})$ models inspired by variable deceleration parameter.}


\author{F. Darabi\altaffilmark{}}
\affil{Department of Physics, Azarbaijan Shahid Madani University , Tabriz, 53714-161 Iran.\\
Research Institute for Astronomy and Astrophysics of Maragha (RIAAM), Maragha 55134-441, Iran}
\email{f.darabi@azaruniv.edu}


%

\begin{abstract}
We study an special law for the deceleration parameter, recently proposed by Akarsu and Dereli, in the context of $f(R)$, $f(T)$ and $f(\mathcal{G})$ theories of modified gravity. This law covers the law of Berman for obtaining exact cosmological models to account for the current acceleration of the universe, and also gives the opportunity to generalize many of the dark energy models having better consistency with the cosmological observations. Our aim is to reconstruct the $f(R)$, $f(T)$ and $f(\mathcal{G})$ models inspired by this law of variable deceleration parameter.  Such models may then exhibit better consistency with the cosmological observations.  
\end{abstract}


\maketitle

\section{Introduction}

It is strongly believed that our universe is now experiencing
an accelerated expansion. This belief is supported by the recent observations from type Ia supernovae \citep{SN,SN1,SN2}, Large Scale Structure
\citep{LSS,LSS1}, and Cosmic Microwave Background anisotropies \citep{CMB}.
Dark energy was the first idea to explain this accelerated expansion in the context of general relativity. This energy is a concept that we usually use for the unknown energy source in general relativity that is believed to be responsible for the observed acceleration of the universe. Modified gravities, on the other hand, are alternative ways for explaining this acceleration. Among the common $f(R)$ modified theories of gravity \citep{19',20',21',Cruz} explaining this acceleration, a theory of scalar-Gauss-Bonnet gravity, so
called $f(\mathcal{G})$ is also proposed \citep{15,15-1,15-2,15-3,15-4,15-5} which is closely related with the low-energy string effective action.
In this proposal, the current acceleration of the universe
is caused by a mixture of scalar phantom and (or) potential/stringy effects.
On the other hand, a theory of $f(T)$ gravity has recently been
received attention. Models based on this modified teleparallel gravity
were presented as an alternative to inflationary
models \citep{16, 17} and dark energy models \citep{18'}. Moreover, attractor solutions in $f(T)$ cosmology \citep{Mubasher3}, generalized second law of thermodynamics in $f(T)$ cosmology with power-law and logarithmic corrected entropies \citep{Mubasher4}, power-law solutions in $f(T)$ gravity \citep{f1}, and cylindrical solutions in modified $f(T)$ gravity \citep{Mubasher5} are among the interesting recent works in this context. A variant of models for modified gravities has also been obtained by using the Noether symmetry approach \citep{Salvatore1,Vakili,Mubasher1,Mubasher2,
Hao,Farhad}.

Recently, Akarsu and Dereli proposed a special law for the deceleration parameter which is linear in time with a negative slope. This law covers the law of Berman \citep{Berman83,Berman88} (where the deceleration
parameter is constant) used for obtaining exact cosmological models,
in the context of dark energy, to account for the current acceleration of the universe. More recently, a comparison of this law with the standard $\Lambda$CDM cosmology has been presented \citep{o-1}, and also LRS Bianchi type-I
cosmological model with linearly varying deceleration parameter has been
studied \citep{Adhav}. According to this law, only the spatially closed and flat universes with cosmological fluid are allowed and the universe ends with a big-rip. In principle, this new law gives the opportunity to generalize
many of these dark energy models having better consistency with the cosmological observations.

The linearly varying deceleration parameter of Akarsu and Dereli is defined by \citep{o}
\begin{equation}\label{1a}
q=-\frac{a\ddot a}{\dot a ^2}=-1-\frac{\dot{H}}{H^2}=-kt+m-1,
\end{equation}
where $a(t)$ and $H(t)$ are time dependent scale factor and Hubble parameter, respectively, $k$ and $m$ are positive constants and an over dot denotes the time derivative. Solving (\ref{1a}) for the scale
factor, we obtain
\begin{eqnarray}
a(t)&=&a_1e^{\frac{2}{m}{\mbox{arctanh}}(\frac{k}{m}t-1)}, \ \ \ k>0, \ \ m>1,\label{1b}\\
a(t)&=&a_2(mt+c_2)^{1/m}, \ \ k=0,\ \ m>0,\label{1c}\\
a(t)&=&a_3e^{c_3t},\ \ k=0, \ \ m=0,\label{1d}
\end{eqnarray}
where $a_1, a_2, a_3$ and $c_2, c_3$ are constants of integration. 
In this paper, we aim to find the $f(R)$, $f(T)$ and $f(\mathcal{G})$ models inspired by the above law of variable deceleration parameter. This is called
{\it reconstruction method}. This method has already been used by people
in different models. In fact, considering the modified $f(R), f(T)$ and $f(\mathcal{G})$
gravity models as effective descriptions of the underlying theory
of dark energy, it is interesting to study how these modified gravities can describe the dark energy. On the other hand, according to the law of variable deceleration parameter, one may take the opportunity to generalize many of these dark energy models with better observational consistency \citep{o}. Therefore, motivated by the above comments on the importance of modified
gravities and the law of variable deceleration parameter, it is very appealing to study how these modified gravities can effectively describe the law of variable deceleration parameter. Reconstruction for $f(R)$ was first developed by Nojiri {\it et al} for gravity consistent with realistic cosmology and universe expansion history \citep{Nojiri1,Nojiri2}. Another variant of reconstruction is developed by Nojiri {\it et al} in terms of e-folding \citep{Nojiri3}.
General review on $f(R)$ gravity, its singularity structure and its 
reconstruction program (as well as $f(\mathcal G)$, etc) is given in \citep{Nojiri4}.
Recently, reconstruction of $f(T)$ gravity is also studied in \citep{Nojiri5}.

In section 2, we study the above three cases in the context of modified $f(R)$ gravity to find the corresponding $f(R)$ models. In section 3, we study these cases in the context of teleparallel equivalent of General Relativity to find the corresponding $f(T)$ models. In section 4, we study these cases in the context of scalar-Gauss-Bonnet gravity to find the corresponding $f(\mathcal{G})$ models. The paper ends with a brief conclusion.     

\section{Variable Deceleration Parameter and $f(R)$ Model}

The most popular theory of modified gravities is the one known as $f(R)$ theory of gravity. The action for this theory coupled with matter 
$S_m$ is given by \citep{19'}, \citep{20',20'-1} and \citep{21'} 
\begin{equation}\label{0}
S=\frac{1}{2\kappa^2}\int d^4x \sqrt{-g} f(R)+S_m,
\end{equation}
where $\kappa^2=8\pi G$. In agreement with the current
observations, we take the spatially-flat Friedmann-Robertson-Walker (FRW)
metric  
\begin {equation}\label{4}
ds^{2}=-dt^{2}+a(t)^{2}\sum^{3}_{i=1}(dx^{i})^{2}.
\end{equation}
Moreover, we assume the matter source to
be a perfect fluid. By using the metric (\ref{4}) and the perfect fluid in the Lagrangian (\ref{0}) we obtain the field equations 
\begin{eqnarray}\label{44}
 -\frac{f(R)}{2} + 3\left(H^2 + \dot{H}\right) f_R(R)
 - 18 \left( 4H^2 \dot{H} + H \ddot{H}\right) f_{RR}(R)
+ \kappa^2 \rho=0\,,
\end{eqnarray}
\begin{eqnarray}\label{Cr4b}
  \frac{f(R)}{2}- \left(\dot{H} + 3H^2\right)f_R(R)
&+& 6 \left( 8H^2 \dot{H}+ 4 {\dot{H}}^2 + 6 H \ddot{H} + d\ddot{H}/dt\right) f_{RR}(R)\\ \nonumber
&+& 36\left( 4H\dot H + \ddot H\right)^2 f_{RRR}(R)  
+ \kappa^2 p=0\,, 
\end{eqnarray}
where $f_R=f'(R)$, $f_{RR}=f''(R)$, $f_{RRR}=f'''(R)$, the Hubble rate $H$ is defined by $H=\dot a/a$ and the scalar curvature $R$ is given by 
\begin{equation}\label{444}
R=12H^2 + 6\dot H.
\end{equation}
Here, $\rho$ and $p$ are the energy density and pressure of the matter source,
respectively. The equation of energy conservation is also obtained as usual
\begin{equation}\label{6}
\dot\rho+3H(\rho+p)=0.
\end{equation}

\subsection*{\Large{Case 1}}

By using Eq.(\ref{1b}) in Eq.(\ref{6}), we obtain
\begin{equation}\label{77}
\rho(t)=C_1\left[\frac{t}{kt-2m}\right]^{-\frac{3(1+w)}{m}},
\end{equation}
where use has been made of the barotropic equation of state $p=w\rho$ with
$w$ being a constant, and $C_1$ is a constant of integration. We then find
\begin{equation}\label{77-0}
H(t)=2\,k{m}^{-2} \left[ 1- \left[ {\frac {kt}{m}}-1 \right] ^{2} \right] ^{-1},
\end{equation}
\begin{equation}\label{77-00}
\dot{H(t)}=4\,k^2{m}^{-3} \left[ 1- \left[ {\frac {kt}{m}}-1 \right] ^{2} \right] ^{-2}\left[ {\frac {kt}{m}}-1 \right].
\end{equation}
Inverting $H(t)$ to obtain the inverse function leads to
\begin{equation}\label{77-1}
t(H)={\frac {Hm\pm\sqrt {{H}^{2}{m}^{2}-2\,Hk}}{Hk}},
\end{equation}
where the positive sign will be selected so that $t>0$. 
Moreover, by using Eq.(\ref{1a}), we easily obtain
\begin{equation}
\dot{H}=-(q+1)H^2.
\end{equation}
Inserting this result in the right hand side of Eq.(\ref{444})
gives the following equation 
\begin{equation}\label{4444}
R=6H^2(1-q) .
\end{equation}
By inserting $t(H)$ in Eq.(\ref{77}), and using Eqs.(\ref{444}), 
(\ref{77-00}) and (\ref{4444}) we get 
\begin{equation}\label{77-2}
\rho(R)=C_1e^{\frac{3(1+\omega)}{m}}k\left[ 1-\frac{2}
{ 1+\sqrt{1-\frac{2k}{m^2\sqrt{R/6(1-q)}}}}\right].
\end{equation}
The Friedmann equation (\ref{44}) takes on the following form

$$
\left[\frac{2R^2(1+q)}{(1-q)^2}-36\left(\frac{(1+q)^{4/3}R}{6(1-q)}\right)^{3/2}+18\dot{q}\frac{R}{6(1-q)}\right]f_{RR}(R)-\frac{f(R)}{2}-\frac{3qR}{6(1-q)}f_{R}(R)\\
$$
\begin{eqnarray}\label{888}
+k^2C_1e^{\frac{3(1+\omega)}{m}}k\left[ 1-\frac{2}
{ 1+\sqrt{1-\frac{2k}{m^2\sqrt{R/6(1-q)}}}}\right]=0
\end{eqnarray}
Note that $q$ can be expressed in terms of $R$ through Eqs.(\ref{77-1}), and (\ref{4444}). Therefore, in principle, the above equation is a differential equation for $f(R)$. There is no known analytical solution for this equation, therefore one can look for numerical or approximate solutions.

\subsection*{\Large{Case 2}}

Now, we use the second case for the scale factor. Putting Eq.(\ref{1c}) in
Eq.(\ref{6}), and using $H=(mt+c_2)^{-1}$ and Eq.(\ref{444}) leads to a simpler
form of $\rho$ as 
$$
\rho(t)=A_1(mt+c_2)^{\frac{-3(1+w)}{m}}=\frac{A_1}{a_2}a^{-3(1+\omega)}=A_1\left(H^2\right)^{\frac{3(1+w)}{2m}},
$$
or
\begin{equation}\label{100}
\rho(R)=A_1\left[\frac{R}{6(1-q)}\right]_,^{\frac{3(1+w)}{2m}}
\end{equation}
where $A_1$ is a constant of integration and use has been made of Eq.(\ref{4444}). Inserting Eq.(\ref{100}) in
Eq.(\ref{44}) leads to the following differential
equation for $f(R)$
$$
\left[\frac{2R^2(1+q)}{(1-q)^2}-36\left(\frac{(1+q)^{4/3}R}{6(1-q)}\right)^{3/2}+18\dot{q}\frac{R}{6(1-q)}\right]f_{RR}(R)-\frac{f(R)}{2}-\frac{3qR}{6(1-q)}f_{R}(R)\\
$$
\begin{eqnarray}\label{8888}
+k^2A_1\left[\frac{R}{6(1-q)}\right]^{\frac{3(1+w)}{2m}}=0.
\end{eqnarray}
As in the previous case, there is no known analytical solution for this equation, so one can look for numerical or approximate solutions.

\subsection*{\Large{Case 3}}

For the third case, we have $H={\rm Const}$
and $R={\rm Const}$. The Friedmann equation (\ref{44}) reads  
\begin{eqnarray}\label{404}
 -\frac{f(R)}{2} + 3H_0^2 f_R(R)
 + \kappa^2 \rho_0=0\,,
\end{eqnarray}
where both $H_0$ and $\rho_0$ are constant. The solution of this equation is simply obtained as
\begin{equation}
f(R)=A_2 \exp{\left(\frac{R}{6H_0^2}\right)}+2k^2\rho_0,
\end{equation}
where $A_2$ is a constant of integration.

\section{Variable Deceleration Parameter and $f(T)$ Model}

The theory of modified gravity based on a modification of the teleparallel equivalent of General Relativity is called $f(T)$ theory of gravity. The action for such a theory coupled with matter 
$L_m$ is given by \citep{18'}, \citep{19} and \citep{20}
\begin{equation}\label{1}
S=\frac{1}{2\kappa^2}\int d^4x~ e~ [T+f(T)]+S_m,
\end{equation}
where $e=det(e^i_{\mu})=\sqrt{-g}$. The teleparallel Lagrangian
$T$ is defined as 
\begin{equation}\label{2}
T=S^{\:\:\:\mu \nu}_{\rho} T_{\:\:\:\mu \nu}^{\rho},
\end{equation}
where
$$
T_{\:\:\:\mu \nu}^{\rho}=e_i^{\rho}(\partial_{\mu}
e^i_{\nu}-\partial_{\nu} e^i_{\mu}),
$$
$$
S^{\:\:\:\mu \nu}_{\rho}=\frac{1}{2}(K^{\mu
\nu}_{\:\:\:\:\:\rho}+\delta^{\mu}_{\rho} T^{\theta
\nu}_{\:\:\:\theta}-\delta^{\nu}_{\rho} T^{\theta
\mu}_{\:\:\:\theta}),
$$
and $K^{\mu \nu}_{\:\:\:\:\:\rho}$ is the contorsion tensor
$$
K^{\mu \nu}_{\:\:\:\:\:\rho}=-\frac{1}{2}(T^{\mu
\nu}_{\:\:\:\:\:\rho}-T^{\nu \mu}_{\:\:\:\:\:\rho}-T^{\:\:\:\mu
\nu}_{\rho}).
$$
The dynamical variable in the teleparallel equivalent of General Relativity is the vierbein $e^i_{\mu}$. The field equations are then obtained 
\begin {equation}\label{4'}
T=-6H^2,
\end{equation}
\begin {equation}\label{5}
H^2=\frac{\kappa^2 \rho}{3}-\frac{1}{6}f-2H^2f_T,
\end{equation}
\begin {equation}\label{5'}
\dot{H}=-\frac{\kappa^2 (\rho+p)}{2(1+f_T-12H^2f_{TT})},
\end{equation}
where $f_T=f'(T)$ and $f_{TT}=f''(T)$. Now, we investigate $f(T)$ for three cases obtained in Eqs.(\ref{1b}), (\ref{1c}), and (\ref{1d}) as follows.

\subsection*{\Large{Case 1}}

In this case, Eqs.(\ref{77}), (\ref{77-0}), (\ref{77-1}) and (\ref{4'}), lead to
\begin{equation}\label{7-2}
\rho(T)=C_2e^{\frac{3(1+\omega)}{m}}k\left[ 1-\frac{2}
{ 1+\sqrt{1-\frac{2k}{m^2\sqrt{-T/6}}}}\right],
\end{equation}
where $C_2$ is a constant of integration and the Friedmann equation (\ref{5}) casts in the following differential equation for $f(T)$
\begin{equation}\label{8}
2Tf_T-f(T)+T+2\kappa^2 C_2e^{\frac{3(1+\omega)}{m}}k\left[ 1-\frac{2}
{ 1+\sqrt{1-\frac{2k}{m^2\sqrt{-T/6}}}}\right]=0.
\end{equation}

Like some previous cases, there is no known analytical solution for this equation, and one can look for numerical or approximate solutions.

\subsection*{\Large{Case 2}}

Putting Eq.(\ref{1c}) in Eq.(\ref{6}), and using $H=(mt+c_2)^{-1}$ and Eq.(\ref{4'}) leads to
\begin{equation}\label{10}
\rho(t)=A_3(mt+c_2)^{\frac{-3(1+w)}{m}}=\frac{A_3}{a_2}a^{-3(1+\omega)}=A_3\left(H^2\right)^{\frac{3(1+w)}{2m}}=
A_3\left(-\frac{T}{6}\right)^{\frac{3(1+w)}{2m}},
\end{equation}
where $A_3$ is a constant of integration. Inserting Eq.(\ref{10}) in
Eq.(\ref{5}) leads to the following differential
equation
\begin{equation}\label{11}
2Tf_T-f(T)+T+2\kappa^2 A_3\left(-\frac{T}{6}\right)^{\frac{3(1+w)}{2m}}=0,
\end{equation}
whose solution is obtained as
\begin{equation}\label{12}
f(T)=-T+B_1\sqrt{T}+A_3\frac{2^{4-\frac{3(1+w)}{2m}}3^{-\frac{3(1+w)}{2m}}\frac{\kappa^2}{8}}{1-2{\frac{3(1+w)}{2m}}}(-T)^{\frac{3(1+w)}{2m}},
\end{equation}
where $B_1$ is another constant of integration.

\subsection*{\Large{Case 3}}

Now, we use the third case for the scale factor. For this case, we have $H={\rm Const}$
and $T={\rm Const}$. So, Eq.(\ref{5}) becomes as follows  
\begin{equation}\label{11-0}
2Tf_T-f(T)+T+2\kappa^2\rho_0=0,
\end{equation} 
where $\rho=\rho_0={\rm Const}$. The solution of this equation is simply obtained as
\begin{equation}
f(T)=A_4\sqrt{T}-T+2\kappa^2\rho_0,
\end{equation}
where $A_4$ is a constant of integration.

\section{Variable Deceleration Parameter and $f(\mathcal{G})$ Model}

Now, we consider the following $f(\mathcal{G})$ action which describes Einstein's gravity coupled with perfect fluid plus a function of the Gauss-Bonnet term \citep{Nojiri1}, \citep{Nojiri2}, \citep{f2}
\begin{equation}\label{16}
S=\frac{1}{2\kappa^2}\int d^4x\sqrt{-g}~[R+f(\mathcal{G})]+S_m
,
\end{equation}
where the Gauss-Bonnet invariant is defined by $\mathcal{G}\equiv
R^2-4R_{\mu\nu}R^{\mu\nu}+R_{\mu\nu\lambda\sigma}R^{\mu\nu\lambda\sigma}.$
The Friedmann equation for a flat FRW background is obtained \citep{f2}
\begin{equation}\label{17}
\frac{-3}{\kappa^2}H^2+\mathcal{G}f_\mathcal{G}-f(\mathcal{G})-24\dot{
\mathcal{G}}H^3f_{\mathcal{G}\mathcal{G}}+\rho=0,
\end{equation}
where $\rho$ satisfies the conservation equation (\ref{6}).

For the FRW metric, the Gauss-Bonnet term and Ricci scalar
take the following forms
\begin{equation}\label{18}
\mathcal{G}=24(\dot H H^2 +H^4),
\end{equation}
\begin{equation}\label{18'}
R=6(\dot H+2H^2).
\end{equation}
By using Eq.(\ref{1a}), we easily obtain
\begin{equation}
\dot{H}=-(q+1)H^2,
\end{equation}
and inserting this in the right hand side of $\mathcal{G}$ in Eq.(\ref{18})
gives the following equations 
\begin{equation}\label{g}
\mathcal{G}=-24 q H^4, 
\end{equation}
\begin{equation}\label{g'}
{\dot\mathcal{G}}=24 H^4 [k+4H q(q+1)].
\end{equation}
Now, by using (\ref{g'}), we can write the Friedmann equation (\ref{17}) in the following form 
\begin{equation}\label{fg}
-\frac{3}{\kappa^2}H^2+\mathcal{G}f_\mathcal{G}-f(\mathcal{G})-(24)^2
H^7[k+4Hq(q+1)]f_{\mathcal{G}\mathcal{G}}+\rho=0.
\end{equation}
In this equation, $\rho$, $q$ and $H$ implicity are functions of $\mathcal{G}$ by following this algorithm: By using the function $H(t)$, one may solve it for $t=t(H)$. Then, it is possible to write $\dot{H}=\dot{H}(H)$, and obtain $H=H(\mathcal{G})$ by using Eq.(\ref{18}).
Thus, since $\rho=\rho(H)$ and $q=q(\dot{H},H)$, then we have $\rho=\rho(\mathcal{G})$
and $q=q(\mathcal{G})$. Therefore, we read the Eq.(\ref{fg}) as a differential equation for $f(\mathcal{G})$ which may be solved analytically or numerically. Now, we analyze the models given by (\ref{1b}), (\ref{1c}) and (\ref{1d}).

\subsection*{\Large{Case 1}}

By using Eqs.(\ref{77}), (\ref{77-0}), (\ref{77-1}), and (\ref{g}) we find
\begin{equation}\label{7-2'}
\rho(\mathcal{G})=C_3e^{\frac{3(1+\omega)}{m}}k\left[ 1-\frac{2}
{ 1+\sqrt{1-\frac{2k}{m^2\left(-\frac{\mathcal{G}}{24q}\right)^{1/4}}}}\right],
\end{equation}
where $C_3$ is a constant of integration. By substituting the above $\rho(\mathcal{G})$, and also $H(\mathcal{G})$ from Eq.(\ref{g}), in Eq.(\ref{fg}) we find the following differential equation
for $f(\mathcal{G})$ 
\begin{eqnarray}\label{fg''}
&-&\frac{3}{\kappa^2}\left(-\frac{\mathcal{G}}{24q}\right)^{1/2}+\mathcal{G}f_\mathcal{G}-f(\mathcal{G})\\
\nonumber
&-&(24)^2\left(-\frac{\mathcal{G}}{24q}\right)^{7/4}\left[k+4\left(-\frac{\mathcal{G}}{24q}\right)^{1/4}q(\mathcal{G})(q(\mathcal{G})+1)\right]f_{\mathcal{G}\mathcal{G}}\\
\nonumber
&+&C_3e^{\frac{3(1+\omega)}{m}}k\left[ 1-\frac{2}
{ 1+\sqrt{1-\frac{2k}{m^2\left(-\frac{\mathcal{G}}{24q}\right)^{1/4}}}}\right]=0.
\end{eqnarray}
No analytical solution for this equation is known, therefore one can look for numerical or approximate solutions.

\subsection*{\Large{Case 2}}

In this case we obtain
\begin{equation} \label{36}
H=\frac{1}{mt+c_2}, 
\end{equation}
whose time derivative becomes
\begin{equation}
\dot{H}=-mH^2.
\end{equation}
Comparing the above result with the following equation derived from (\ref{18}),
namely
\begin{equation}\label{h'}
\dot{H}=\frac{\mathcal{G}}{24H^2}-H^2,
\end{equation}
leads to the fourth order equation for $H=H(\mathcal{G})$ as follows
\begin{equation}\label{h''}
H^4-\frac{\mathcal{G}}{24(1-m)}=0.
\end{equation}
Solving this fourth order equation is easy and gives the following solution
\begin{equation}\label{hg2}
H(\mathcal{G})=\pm\left[\frac{\mathcal{G}}{24(1-m)}\right]^{1/4},
\end{equation}
or
\begin{equation}\label{hg2'}
H^2(\mathcal{G})=\left[\frac{\mathcal{G}}{24(1-m)}\right]^{1/2}.
\end{equation}
It is easy to obtain the following expression for $\rho$
\begin{equation}\label{rho}
\rho=\rho_0 (a_2 c_2^{1/m})^{-3(1+w)}(1-m H)^{\frac{3(1 +w)}{m}},
\end{equation}
and by inserting $\rho$ in Eq.(\ref{fg}), we obtain the following differential equation for $f(\mathcal{G})$
\begin{eqnarray}\label{fg2}
&-&\frac{3}{\kappa^2}H^2(\mathcal{G})+\mathcal{G}f_\mathcal{G}-f(\mathcal{G})\\ \nonumber
&-&(24)^2H^7(\mathcal{G})[4H(\mathcal{G})m(m-1)]f_{\mathcal{G}\mathcal{G}}\\ \nonumber
&+&\rho_0(a_2 c_2^{1/m})^{-3(1+w)}[1-m H(\mathcal{G})]^{\frac{3(1 +w)}{m}}=0.
\end{eqnarray}
Now, depending on the two choices for $H(\mathcal{G})$ in (\ref{hg2}),
the solutions of this differential equation can be obtained
for $f(\mathcal{G})$.

\subsection*{\Large{Case 3}}

For this case, we obtain $H={\rm Const}$ and so $a(t)=a_3e^{Ht}$. We may proceed
as follows
\begin{equation}\label{h-0}
H={\rm Const}.\ \ \  \Longrightarrow \:\:\:
0=\dot{H}=\frac{\mathcal{G}}{24H^2}-H^2,
\end{equation}
which results in the following fourth order equation for $H=H(\mathcal{G})$
\begin{equation}\label{h-1}
H^4-\frac{\mathcal{G}}{24}=0,
\end{equation}
with the solution
\begin{equation}\label{h-2}
H(\mathcal{G})=\pm\left(\frac{\mathcal{G}}{24}\right)^{1/4}.
\end{equation}
For energy density we find  $\rho=\rho_0={\rm Const}$, so by inserting $H(\mathcal{G})$ and $\rho$ in Eq.(\ref{fg}) we obtain the following differential equation for $f(\mathcal{G})$
\begin{eqnarray}\label{fg2'}
-\frac{3}{\kappa^2}\left(\frac{\mathcal{G}^3}{24}\right)^{1/2}+\mathcal{G}f_\mathcal{G}-f(\mathcal{G}) +\rho_0=0.
\end{eqnarray}
This equation is easily solved and we obtain the following solution
\begin{eqnarray}
f(\mathcal{G}) =\frac{\sqrt{3}}{\sqrt{2}\kappa^2}\mathcal{G}\sqrt{\mathcal{G}}+C_4\mathcal{G}+\rho_0,
\end{eqnarray}
where $C_4$ is the constant of integration.

\section{Final Remarks}

We have studied the recently proposed special law for the deceleration parameter
by Akarsu and Dereli. According to this law only the spatially
closed and flat universes are allowed and in both cases the cosmological fluid exhibits quintom like behavior, moreover the universe ends with a big-rip. This new law also gives the opportunity to generalize many of these dark energy models with better observational consistency. Motivated by this claim,
we have tried to obtain the models of $f(R)$, $f(T)$ and $f(\mathcal{G})$ corresponding to this law in the framework of dark energy models. In this regard, we have obtained some exact solutions, and in some complicate cases we have left the solutions for numerical analysis. In a parallel way to that of Akarsu and Dereli, these solutions may give the opportunity to generalize many of the modified gravity models with better observational consistency.

\section*{Acknowledgment}
I would like to thank the anonymous referee for enlightening and very valuable comments on this paper. This work has been supported financially by Research Institute for Astronomy and Astrophysics of Maragha (RIAAM) under research project No.1/2782-4.

\nocite{*}
\bibliographystyle{spr-mp-nameyear-cnd}
\bibliography{biblio-u1}

\begin{thebibliography}{}

\bibitem[Abazajian et al. (2004)]{LSS}Abazajian~ K.~{\it et al.}  
[SDSS Collaboration], 2004, Astron.\ J.\ 128, 502.
\bibitem[Abazajian et al. (2005)]{LSS1}Abazajian~ K.~{\it et al.}  
[SDSS Collaboration], 2005, Astron.\ J.\ 129, 1755.
\bibitem[Adhav (2011)]{Adhav}Adhav. K. S., 2011, EPJ Plus 126, 122.
\bibitem[Akarsu and Dereli (2012)]{o}Akarsu. O., Dereli. T., 2012, Int. J. Theor. Phys. 51, 612.
\bibitem[Akarsu and Dereli (2012)]{o-1}Akarsu. O, Dereli. T., 2012, Int. J. Theor. Phys, DOI 10.1007/s10773-012-1200-0.
\bibitem[Astier et al. (2006)]{SN2}Astier~P.~{\it et al.}, 2006, Astron.\ Astrophys.\ 447, 31.
\bibitem[Atazadeh and Darabi (2012)]{Farhad}Atazadeh. K, Darabi. F, 2012,
Eur. Phys. J. C 72, 2016.
\bibitem[Bamba et al. (2012)]{Nojiri5}Bamba.  K., Myrzakulov. R., Nojiri.
S., and Odintsov. S. D., 2012, Phys. Rev. D 85, 104036.
\bibitem[Bamba et al. (2012)]{Mubasher4}Bamba. K, Jamil. M, Momeni. D, Myrzakulov.
R, {\it Generalized Second Law of Thermodynamics in $f(T)$ Cosmology with Power-Law and Logarithmic Corrected Entropies}, 2012, arXiv:1202.6114.
\bibitem[Bengochea and Ferraro (2009)]{18'}Bengochea. G, and Ferraro.R, 2009, Phys. Rev. D 79, 124019.
\bibitem[Berman (1983)]{Berman83}Berman. M. S., 1983, Nuovo Cimento B 74, 182.
\bibitem[Berman and Gomide (1988)]{Berman88}Berman. M. S., Gomide. F. M., 1988, Gen. Rel. Grav. 20, 191.
\bibitem[Cai et al. (2011)]{20}Cai. Y-Fu, Chen. S-Hung, Dent. J. B., Dutta.
S., and Saridakis. E. N., 2011, Class. Quantum Grav. 28, 215011.
\bibitem[Capozziello and De Felice (2008)]{Salvatore1}Capozziello. S, De Felice. A, 2008, J. Cosmol. Astropart. Phys. 0808, 016.
\bibitem[Capozziello et al. (2009)]{20'}Capozziello. S., De Laurentis. M., and  Faraoni. V, 2009, arXiv:0909.4672.
\bibitem[Capozziello and Francaviglia (2008)]{20'-1}Capozziello. S, and Francaviglia. M., 2008, Gen. Rel. Grav. 40, 357.
\bibitem[Carter and Neupane I (2006)]{15-2}Carter. B. M. N., and Neupane. I. P., 2006, Phys. Lett. B 638, 94.
\bibitem[Carter and Neupane II (2006)]{15-3}Carter. B. M. N., and Neupane. I. P., 2006, JCAP 0606, 004.
\bibitem[Cruz and Dobado (2006)]{Cruz}Cruz-Dombriz. A. de la, Dobado. A., 2006, Phys. Rev. D 74, 087501.
\bibitem[Elizalde et al. (2010)]{15-4}Elizalde. E., Myrzakulov. R., Obukhov. V. V., Sáez-Gómez. D., 2010, Class. Quant. Grav. 27, 095007. 
\bibitem[Ferraro and Fiorini (2007)]{16}Ferraro. R., and Fiorini. F., 2007,
Phys. Rev. D 75, 084031.
\bibitem[Ferraro and Fiorini (2008)]{17}Ferraro. R., and Fiorini. F., 2008, Phys. Rev. D 78, 124019.
\bibitem[Houndjo et al. (2012)]{Mubasher5}Houndjo. M. J. S., Momeni. D, Myrzakulov.
R, {\it Cylindrical Solutions in Modified f(T) Gravity}, 2012, arXiv:1206.3938.
\bibitem[Hussain et al. (2012)]{Mubasher1}Hussain. I, Jamil. M, Mahomed.
F. M., 2012, Astrophys. Space. Sci. 337, 373.
\bibitem[Jamil et al. (2011)]{Mubasher2}Jamil. M, Mahomed. F. M., Momeni.
D, 2011, Phys. Lett. B 702, 315.
\bibitem[Jamil et al. (2012)]{Mubasher3}Jamil. M, Momeni. D, Myrzakulov.
R, 2012, Eur. Phys. J. C. 72, 1959.
\bibitem[Linder (2010)]{19}Linder. E. V., 2010, Phys. Rev. D 81, 127301.
\bibitem[Myrzakulov et al. (2011)]{15-5}Myrzakulov. R., Sáez-Gómez. D., Tureanu. A., 2011, Gen. Rel. Grav. 43, 1671. 
\bibitem[Nojiri et al. (2005)]{15}Nojiri. S., Odintsov. S. D., and Sasaki. M., 2005, Phys. Rev. D 71, 123509.
\bibitem[Nojiri et al. (2006)]{15-1}Nojiri. S., Odintsov. S. D., and Sami. M., 2006, Phys. Rev. D 74, 046004.
\bibitem[Nojiri and Odintsov (2007)]{19'}Nojiri. S., Odintsov. S. D., 2007, Int. J. Geom. Meth. Mod. Phys. 4, 115.
\bibitem[Nojiri and Odintsov (2005)]{Nojiri1}Nojiri. S., Odintsov. S. D., 2005, Phys. Lett. B 631, 1.
\bibitem[Nojiri et al. (2006)]{Nojiri2}Nojiri. S., Odintsov. S. D., Gorbunova. O. G., 2006, J. Phys. A 39, 6627.
\bibitem[Nojiri et al. (2009)]{Nojiri3}Nojiri. S., Odintsov. S. D., and Saez-Gomez. D., {\it Cosmological reconstruction of realistic modified $f(R)$ gravities} arXiv:0908.1269.
\bibitem[Nojiri and Odintsov (2011)]{Nojiri4}Nojiri. S., Odintsov. S. D., 2011, Phys. Rept. 505, 59.
\bibitem[Perlmutter et al. (1999)]{SN1}Perlmutter~S.~{\it et al.} [Supernova Cosmology Project Collaboration], 1999, Astrophys.\ J.\ 517, 565.
\bibitem[Rastkar et al. (2012)]{f2}Rastkar. A. R., Setare. M. R., and Darabi.
F., 2012, Astrophys Space Sci 337, 487.
\bibitem[Riess et al. (1998)]{SN}Riess~A.~G.~ {\it et al.} [Supernova Search Team Collaboration], 1998, Astron.\ J.\ 116, 1009. 
\bibitem[Setare and Darabi (2012)]{f1} Setare. M. R., Darabi. F., 2012, To
appear in Gen. Rel. Grav. arXiv:1110.3962v1.
\bibitem[Sotiriou and Faraoni (2010)]{21'}Sotiriou. T. P.,  and  Faraoni.
V., 2010, Rev. Mod. Phys. 82, 451.
\bibitem[Spergel (2003)]{CMB}Spergel~D.~N.~{\it et al.}  
[WMAP Collaboration], 2003, Astrophys.\ J.\ Suppl.\ 148, 175.
\bibitem[Vakili (2008)]{Vakili}Vakili. B, 2008, Phys. Lett. B 664, 16.
\bibitem[Wei et al. (2012)]{Hao}Wei. H, Guo. X-J., Wang. L-F, 2012, Phys. Lett. B 707, 298. 





\end{thebibliography}

\end{document}